\documentclass[10pt]{article}
\usepackage{graphicx}

\begin{document}

\title{Spectropolarimetry of the Deep Impact Target Comet 9P/Tempel 1 with HiVIS}
\author{D. M. Harrington$^1$, K. Meech$^1$, L. Kolokolova$^2$, J.R. Kuhn$^1$, and K. Whitman$^1$}
\date{July 27 2006}
\maketitle

Email:  dmh@ifa.hawaii.edu

$^1$  University of Hawaii, Honolulu, HI  96822

$^2$  University of Maryland, College Park MD 20742

\section{Abstract}

High resolution spectropolarimetry of the Deep Impact target, comet 9P/ Tempel 1, was performed during the impact event on July 4th, 2005 with the HiVIS Spectropolarimeter and the AEOS 3.67m telescope on Haleakala, Maui. We observed atypical polarization spectra that changed significantly in the few hours after the impact.  The polarization of scattered light as a function of wavelength is very sensitive to the size and composition (complex refractive index) of the scattering particles as well as the scattering geometry.   As opposed to most observations of cometary dust, which show an increase in the linear  polarization with the wavelength (at least in the visible domain and  for phase angles greater than about 30¡, a red polarization  spectrum) observations of 9P/Tempel 1 at a phase angle of 41$^\circ$ beginning 8 minutes after impact and centered  at 6:30UT showed a polarization of 4\% at 650 nm falling to 3\% at 950 nm.  The  next observation, centered an hour later showed a polarization of  7\% at 650 nm falling to 2\% at 950nm.  This corresponds to a spectropolarimetric gradient, or slope, of -0.9\% per 1000\AA\ 40 minutes after impact,  decreasing to a slope of -2.3\% per 1000\AA\ an hour and a half after impact.   This is an atypical blue polarization slope, which became more blue 1 hour after impact.   The polarization values of 4\% and 7\% at  650nm are typical for comets at this scattering angle,  whereas the low polarization of 2\% and 3\% at 950nm is not.  We compare observations of  comet 9P/Tempel 1 to that of a typical comet, C/2004 Machholz, at a phase angle of 30$^\circ$ which showed a typical red slope, rising from 2\% at 650nm to 3\% at 950nm in two different observations (+1.0 and +0.9\% per 1000\AA).

Keywords: Polarimetry, Comets, Individual (9P/Tempel 1), Comets, Composition, Impact Processes

\section{Introduction}
	
The scattered light from cometary coma is linearly polarized. The degree of polarization can reach 10's of percent and is measured as a function of the scattering geometry (phase angle) and wavelength.  Angular and spectral dependence of polarization are signatures of the size, composition and structure of comet grains (Hanner 2002, Hadamcik \& Levasseur-Regourd 2003c, Kolokolova {\it et al.} 2004 and references therein, Kimura {\it et al.} 2006, Hadamcik {\it et. al.} 2006).

Comets typically show a degree of polarization that increases with wavelength. This is called a red polarimetric color for the dust coma.  The values of polarimetric color depend on the phase angle (typically increasing with phase angle), reaching $\sim$0.8\% per 1000\AA\ at phase angle $\sim95^\circ$, the phase angles of maximum polarization (Kolokolova {\it et al.} 2004, Kolokolova and Jockers 1997, Levasseur-Regourd \& Hadamcik 2003).  Few comets have shown a blue polarimetric color, {\it i.e.} a decrease of polarization with wavelength, such as comet 21P/Giacobini-Zinner falling from 8\% at 443nm to 5\% at 642nm (Kiselev {\it et al.} 2000). The typical red polarization of cometary dust is a result of the scattering  size-parameter ($2\pi a/\lambda$, where a is the particle radius and $\lambda$ is the wavelength) decreasing with increasing wavelength {\i.e.} the particles are effectively smaller with longer wavelengths.  Small size-parameter scatterers approach the Rayleigh scattering limit where the polarization is strongest (100\% at 90$^\circ$).  If the scatterers are too large to be in the Rayleigh scattering regime, or if there are significant geometrical and compositional effects (e.g. roughness, shape, inhomogeneity, alignment) the polarization may not show typical wavelength and geometrical dependences (see Bohren \& Huffman 1983 and Mishchenko {\it et. al.} 2000).

There are few cometary spectropolarimetric measurements to date, and long-slit high-resolution spectropolarimetry of comets, to our knowledge, has not been published. Most data on the wavelength dependence of polarization are based on the observations done with imaging or aperture instruments using narrow-band filters to avoid contamination from gas emission lines.  High-resolution spectropolarimeters, however, can resolve gas emission lines and allow us to remove this contamination. A few have reported low-resolution (R$\sim$ 100) spectropolarimetry (e.g. Myers \& Nordsieck 1984). 

On 2005 July 4 at 5:52:02 UT comet 9P/Tempel 1 was impacted by a 364 kg spacecraft as part of the NASA Deep Impact mission (A'Hearn {\it et al.} 2005).  The event presented a unique opportunity to compare spectropolarimetric measurements with other measurements performed simultaneously to further constrain the dust properties and gain insight into the models used to simulate the changes of polarization with wavelength, geometry, and dust properties.  

In section 3 we discuss a novel spectropolarimeter mounted at the AEOS telescope, observation procedures and calibrations with this instrument, and then use of this instrument to observe the Deep Impact event.  In section 4 we discuss typical polarization properties of comets and theoretical modeling which allows to extract properties of comet dust from the observations.  Due to the faintness of comet 9P/Tempel 1, we were only able to perform spectropolarimetric measurements twice on the night of impact, centered 40 and 100 minutes after impact.  We will show in section 5 that post-impact spectropolarimetry of comet 9P/Tempel 1 exhibited unusual spectral behavior of polarization that some hours later was observed returned to the normal (Mumma {\it et al.} 2005, Sugita {\it et al.} 2005, Harker {\it et. al. } 2005).  We compare these results to observations of a typical dust-rich comet, C/2004 Machholz in section 6.  The possible reasons for the unusual behavior of comet 9P/Tempel 1 after the impact are discussed in section 7.

\section{The AEOS HiVIS Spectrograph}

The AEOS telescope is a 3.67m, altitude-azimuth telescope.  The HiVIS spectrograph is a cross-dispersed echelle spectrograph using the f/200 coud\`e optical path, with seven reflections (five at $\sim45^\circ$ incidence) before coming into the optics room (Thornton 2002, Thornton {\it et al.} 2003).  Since non-normal incidence reflections change the polarization state of the incident light, a careful calibration of the telescope has been performed as a function of altitude, azimuth and wavelength.  For more detail see Harrington {\it et al.} 2006.  The plate scale is 0.16$''$ pix$^{-1}$. The cross disperser's Red setting was used for all observations (nominally 637.5-968.0 nm).  The spectrograph uses two 2K$\times$4K Lincoln Labs high resistivity CCDs mounted side by side with a 15 pixel gap to form a 4K$\times$4K array.  The array was rebinned 2$\times$2 so that the each output images was 1024$\times$2048 pixels. We used a 1.5$''$$\times$7$''$ slit (970$\times$4530 km at a geocentric distance of $\Delta$=0.89 AU) for all comet observations.

In all of the results presented in this paper, there are occasional gaps in the wavelength coverage which arose from the loss of half of focal plane array resulting from a failure of one of the two CCDs.

\subsection{The Spectropolarimetry Module}

The spectropolarimetry module for the AEOS spectrograph consists of a rotating achromatic half-wave plate and a calcite Savart Plate. The Savart plate consists of two bonded, crossed calcite crystals which separate an incoming beam into two parallel, orthogonally polarized beams displaced in the spatial direction (along the long axis of the slit) at the image plane.  The small deviation in the image plane splits each spectral order detected on the CCD into two orthogonally polarized orders from which orthogonally polarized spectra are extracted separately.  For example, the red cross disperser setting originally shows 19 orders across the CCD, but after insertion of the Savart plate, it shows 38 (2 polarizations for 19 orders). This design contrasts with other spectropolarimeters which use Fresnel rhomb retarders and a Wollaston prism on fiber-fed spectrographs (Donati {\it et al.} 1999, Goodrich \& Cohen 2003, Petit {\it et al.} 2003, Thornton {\it et al.} 2003). Our continuum polarization measurements are stable because the optical path is fixed.

\subsection{Observing With AEOS -- Measuring Linear Polarization}

The Stokes parameters: I, Q and U fully specifiy the linear polarization of light (Collet 1992, Hapke 1993).  Stokes parameter I is the total intensity of the beam.  Stokes parameter Q is the difference between two orthogonal polarization states  (e.g. horizontal and vertical).  Stokes parameter U is the difference between two orthogonal polarization states rotated 45$^\circ$ with respect to Q (e.g. +45$^\circ$ and -45$^\circ$).  These are measured with HiVIS by taking the fractional difference between two polarized spectra.

The observing sequence is to take spectra at 0, 22.5, 45, and 67.5 degrees rotation of the half-wave plate.  A 45$^\circ$ rotation of the waveplate swaps the location of polarized light in the two polarized spectra, moving it from one polarized order to the other on the CCD allowing for cancellation of systematic errors, such as misalignments or chip defects (see Harrington {\it et al.} 2006 for details).

The linear polarization measurements in this paper will be defined as follows:  Let $a$ and $b$ be the two orthogonally polarized spectra from each spectral order with the waveplate at 0$^\circ$.  Let $c$ and $d$ be the two orthogonally polarized spectra with the waveplate at 45$^\circ$. Since a 45$^\circ$ rotation of the waveplate will swap the position of the polarized light incident along the original waveplate axis (say $\pm Q$ on the CCD), but not the unpolarized light or linearly polarized light incident at 22.5$^\circ$ or 67.5$^\circ$ ($\pm$U), a measurement of a normalized $q$ can be defined as:

\begin{equation}
q=\frac{Q}{I}=\frac{1}{2}\left(\frac{a-b}{a+b} - \frac{c-d}{c+d}\right)=\frac{1}{2}(q_{0^\circ}+q_{45^\circ})
\end{equation}

Measurement of Stokes parameter U is the same formula with the waveplate at 22.5$^\circ$ and 67.5$^\circ$.  Since this is a fractional quantity, any correction applied equally to each order, such as spectrophotometric calibrations, vignetting corrections or skyline subtractions, will not change the polarimetry in any significant way.  That is to say that no standard flux calibrations affect the spectropolarimetry.

The simultaneous imaging of orthogonal polarization states also allows for a more efficient observing sequence and for reduction of systematic effects.  Since both polarized spectra are imaged simultaneously, with identical systematic polarization, measurement (Q or U) can be done with a single image, and consistency between images is easy to quantify. Using the fractional polarizations defined above, we can calculate the degree of polarization (P) and the position angle (PA) as follows:

\begin{equation}
P = \sqrt{q^2 + u^2}
\end{equation}

\begin{equation}
PA = \frac{1}{2}tan^{-1}\frac{q}{u}
\end{equation}

A complication arose because the image rotator was not used
so that the position angle of the slit on the sky was a function
of pointing.  This simplified the polarization calibration by fixing 3
non-normal-incidence reflections which induce instrumental polarization as
a function of time and wavelength.  Fixing the rotator mirrors in-place
makes interpretation of the position angle of the plane polarization
difficult because of the overall rotation of the slit with time
(Harrington {\it et al.} 2006).

\begin{figure}[!h,!t,!b]
\includegraphics[width=\linewidth]{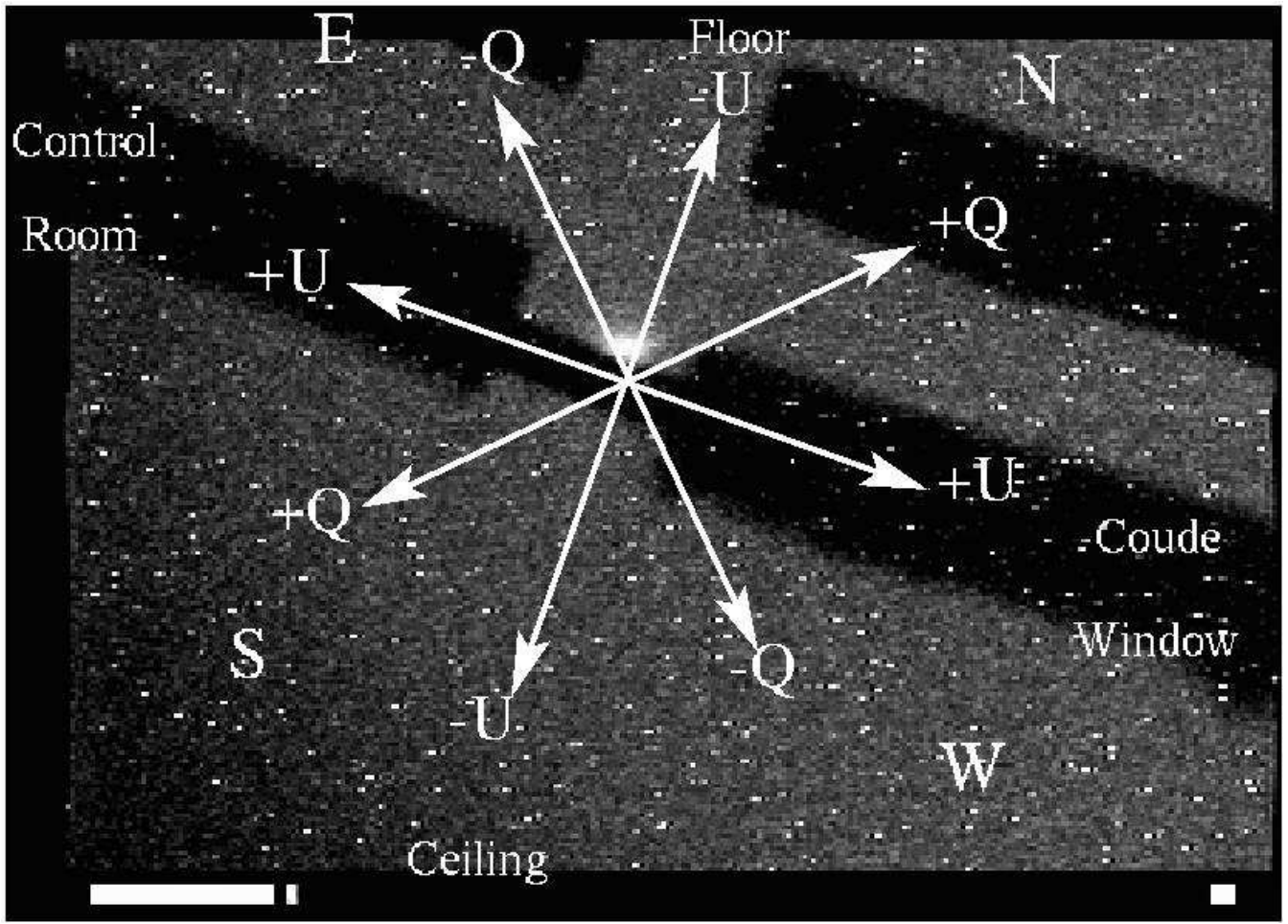}
\caption[SlitQU]{\label{fig-SlitQU}The orientation of Q and U and the projection onto the sky as seen by the
slitviewer camera at altitude=50$^\circ$ azimuth=225$^\circ$.  This is
a rough guide for the orientation of the slit for the impact-night
observations, but the position angle rotates with pointing (by
$\sim10^\circ$) for the comet 9P/Tempel 1 observations.
}
\end{figure}

A frame from the slit-viewer camera is shown in figure \ref{fig-SlitQU}
with the general geometry of the slit and room as seen by the slit
viewer camera.  Stokes U is parallel to the slit, and Q is rotated
$45^\circ$ counter clockwise, pointing to the upper right.  Since the
image rotator was not used, the projection of the slit on the sky is a
function of pointing and is not constant.  We used an optical design
of the telescope to estimate the slit projection onto the sky in the
middle of our observations (7UT), and have plotted this projection (NESW)
in Fig.  \ref{fig-SlitQU}.

A common definition of the degree of polarization used in planetary
science and in many of the references is that the degree of 
polarization is the difference between the
intensity polarized parallel and perpendicular to the scattering plane,
defined below (Kolokolova {\it et al.} 2004):

\begin{equation} 
P_{comet} =
\frac{I_{\perp}-I_{\parallel}}{{I_{\perp}+I_{\parallel}}} 
\end{equation}

Since the image rotator was not used, projection of the slit's position angle onto the sky was not constant, and knowledge of the scattering plane is difficult to quantify.  We will be using the first definition in this paper, since it does not require knowledge of the scattering plane.

Absolute polarization measurements at all wavelengths requires very careful calibration of the telescope.  Since AEOS is an altitude-azimuth telescope, the slit projection and the mirror positions change as the telescope tracks, inducing systematic polarization effects.  Internal optics can induce polarization in the beam but can also depolarize the light or rotate the plane of polarization (Q $\rightarrow$ U).  All of these are functions of wavelength, altitude, and azimuth.  The calibration of the telescope was done by measuring unpolarized standard stars and scattered sunlight as a polarized source, as described in Harrington {\it et al.} (2006).   The first-order correction applied to the spectropolarimetry presented here is a measurement of unpolarized standards to quantify the telescope-induced polarization in unpolarized light.

\section{Polarization of Cometary Dust -- Models and Observations}

	Typically, the degree of polarization of cometary coma is measured using filter polarimeters (aperture and imaging) in the 500-1000nm wavelength range. The polarization is usually  0 to 30\% with a strong dependence on phase angle (scattering geometry) and a more moderate dependence on wavelength.  Kolokolova {\it et al.} (2004) provides a detailed review.  The polarization is small (0-3\%) and parallel to the scattering plane (negative polarization) for phase angles less than 21$^\circ$.  The polarization becomes perpendicular to the scattering plane for phase angles greater than 21$^\circ$ and rises with increasing phase angle until a maximum polarization of 10-30\% is reached at phase angles of 90-100$^\circ$.    
	
	The polarization also shows a wavelength dependence, usually increasing towards 1$\mu$m and flattening off in the IR (Levasseur-Regourd \& Hadamcik 2003).  Some exceptions are known however.  Kiselev (2000) reports a blue polarization spectrum for comet 21P/Giacobini-Zinner (8\% at 443nm falling to 5\% at 642nm at a phase angle of 44$^\circ$).  A blue polarization color was revealed by Giotto in-situ observations of comet 1P/Halley for distances from the nucleus less than 2000 km, and suspected in observations of C/1995 O1 Hale-Bopp (Levasseur-Regourd {\it et al.} 1999, Levasseur-Regourd \& Hadamcik 2003).  

	Some dependence on comet activity was also found and explained as a consequence of the influence of scattering particles of differing size and composition as well as unpolarized emission from gas which contaminated the observations (Hadamcik \& Levasseur-Regourd 2003a, b, c).  A change in polarization due to a change of composition and size is also seen in models (Kolokolova \& Jockers 1997, Kolokolova {\it et al.} 2004, Kimura {\it et al.} 2006, Lasue \& Levasseur-Regourd 2006).

	During the Deep Impact encounter, comet 9P/Tempel 1 was at a phase angle of $\sim$40.9$^\circ$ where comets typically show 4\%-10\% percent polarization in the red filter, integrated across the coma (Kolokolova {\it et al.} 2004).  
 
	Details of the angular and wavelength dependence of linear polarization can provide information about the dust particles: their size, composition, structure and can be used to constrain the particle characteristics from the observations. For this purpose, light scattering by model particles has been studied theoretically and experimentally in great depth ({\it e.g.} Gustafson \& Kolokolova 1999, Kolokolova {\it et al.} 2004, Kimura {\it et al.} 2006, Lasue \& Levasseur-Regourd 2006) over a great range of variables (size, composition, scattering geometry, particle roughness, aggregation processes).  Models that fit the typical angle and red wavelength dependence of the polarization best use aggregates of 10-1000 monomers having small radii ($\sim$100nm).  Model aggregates that fit the typical polarization measurements have a high index of refraction (m=n + ik, n=1.8 to 2.0, k$\sim$0.4 to 0.6).  This corresponds to volume fractions of one-third silicates, two-thirds carbonaceous materials, and a small amount of iron-bearing sulfides (Kimura {\it et al.} 2006).  This material is quite dark (high complex index of refraction) and similar to the composition of 1P/Halley's dust  (Mann {\it et al.} 2004).  The carbonaceous material in these models is roughly two thirds amorphous carbon and one third organic-refractory material.  The type of aggregation (cluster-cluster vs. particle-cluster) did not play a major role in the polarization models, but it does influence the IR spectra (Kimura \& Mann 2004).  Blue polarization gradients were not extensively modeled because of their rarity, but some models demonstrated the tendency of polarimetric color to decrease, {\it i.e.} get more blue, with increasing size, albedo ({\it e.g.} icy), and transparency of the material, whereas more absorbing materials typically have red polarization (Kimura {\it et al.} 2006).  Multiple scattering in optically thick dust clouds depolarizes the scattered light.
	
	With all this information contained in the polarization of scattered light, we expected that the polarimetric data, especially the wavelength dependence of polarization, can shed light on the change in the properties of the dust in comet 9P/Tempel 1 resulting from the Deep Impact encounter.

\section{Deep Impact Spectropolarimetry}

Observations of comet 9P/Tempel 1 were performed on June 30th, July 1st and July 3rd-6th totaling 8.5 hours of integration.  Unfortunately, the comet was  bright enough to obtain useful spectropolarimetry only on the night of impact.  A 60 minute exposure on the pre- and post-impact nights gave as many counts as a 15 minute exposure on impact night. Since spectropolarimetry with HiVIS requires a minimum of 2 exposures to measure Stokes Q and U, and canceling systematic errors requires 4 exposures, spectropolarimetry was only performed on the eight 15-minute exposures starting at 6 UT (8 minutes after impact).  These eight exposures make two complete data sets, from 6-7 and 7-8 UT.  The journal of observations is shown in Table \ref{tab-obs}.  The data were reduced using the AEOS-specific reduction pipelines developed for this instrument (Harrington {\it et al.} 2006).

\begin{figure}[!h,!t,!b]
\includegraphics[width=1.0\linewidth]{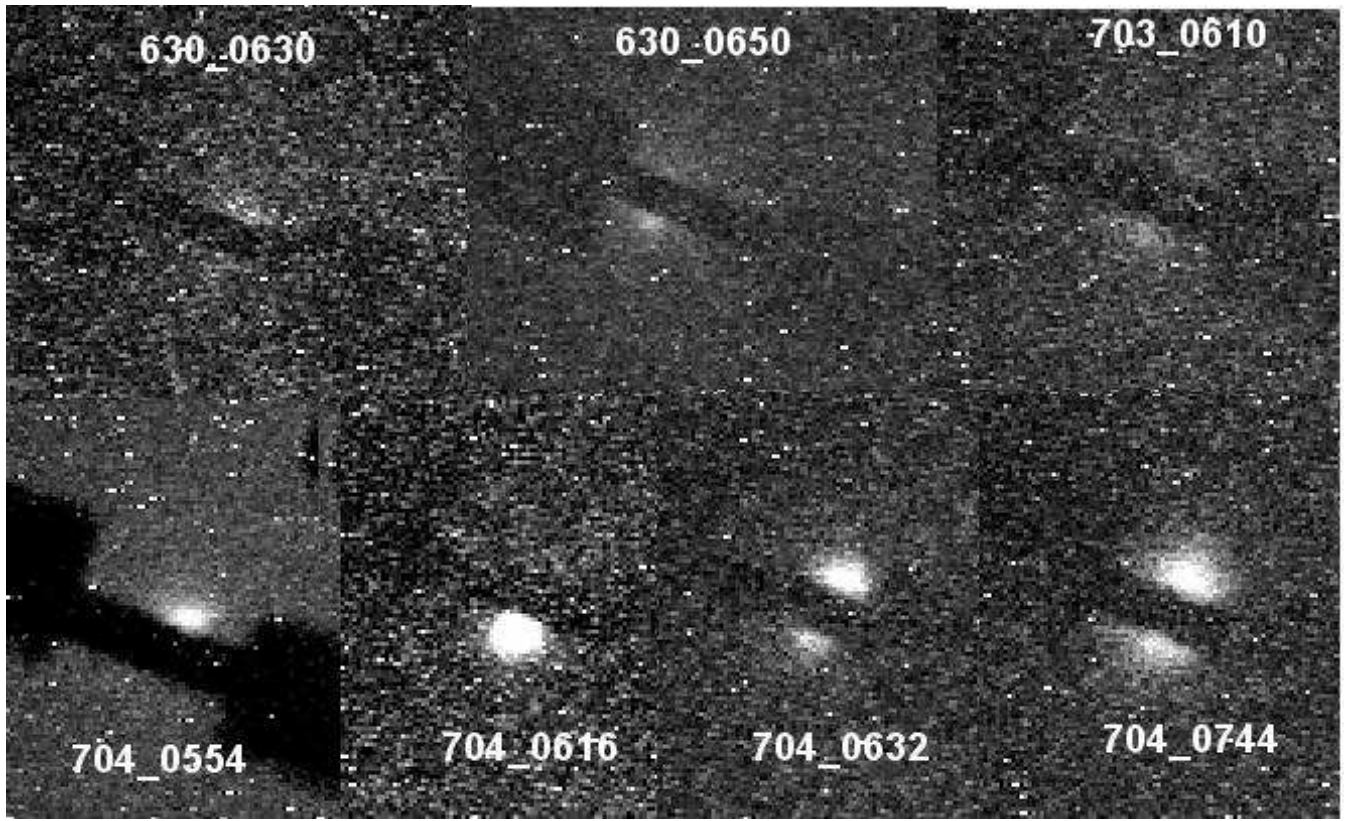}
\caption[slitview]{\label{fig-slitview}Slitviewer camera images -- The date and time of each 1-minute exposure
are included in mdd\_hhmm format.  Three pre-impact frames are shown on
top and four impact-night frames on bottom.  The impact frames show the
ejecta evolution at times 5:54, 6:16, 6:32 and 7:44 UT.  Note that AEOS
non-siderial guiding reset near the start of each exposure, causing a
slight offset of the comet in these slitviewer frames.  The comet was
recentered within 30 seconds, causing minimal losses.
}
\end{figure}

The impact event produced a $\sim$four-fold increase in our observed counts over the night and can be seen as an increase in brightness and a spreading of the ejecta in the slit-viewer camera frames tiled in Fig. \ref{fig-slitview}.  A 5-min exposure started 30 seconds before impact did not detect the comet. Three faint, pre-impact slit-viewer frames from late June are shown in the top row and four frames from the night of impact are shown in the bottom row.  Notice how the brightness increased dramatically and how the comet became more diffuse with time. Since the slit is 1.5$''$ wide and comet 9P/Tempel 1 was at $\Delta$=0.89 AU, a body moving at 1 km sec$^{-1}$ could move across the slit during the 900 sec exposure.  Meech {\it et al.} (2005) report that the leading edge of the ejecta plume was moving out from the nuclues at $\sim$200m s$^{-1}$, giving a minimum slit crossing time of 40 minutes.  This sets the timescale for significant change in scattering ejected gas and dust.  The target became visibly brighter in our slit-viewer 5 minutes after the impact.  The coma continued to increase significantly in size and intensity from exposure to exposure until the comet set.  Fig. \ref{fig-slitview} shows the evolution of the  ejecta as seen on our slit-viewer camera. Since AEOS is designed to  track objects very quickly, non-sidereal guiding is done with ephemeris files.  We used ephemeris files of 900 sec duration, corresponding to our exposure times. Resetting the ephemeris files during readout causes a small change in pointing near the start of each exposure, causing the comet offset seen in Fig. \ref{fig-slitview}.  The guiding was recentered within 30 seconds, causing minimal effect on the observations.

The observed behavior of the comet is consistent with other photometric
and spectroscopic observations (Jehin {\it et al.} 2006, Meech {\it
et al.} 2005, Schleicher {\it et al.} 2005, Sugita {\it et al.} 2005).
Sugita {\it et al.}(2005) reported a plume opening at a position-angle of
225$^\circ$ reaching a size of $\sim$1.5$''$ by 7:10 UT.  Schleicher {\it
et al.} (2006) reported the V and R band fluxes increasing 6-fold by
6:30 UT with a change in behavior in overall flux increase and between
individual V and R band fluxes occuring near 6:30 UT.  Jehin {\it et
al.} (2006) reported the 388nm dust continuum increasing by a factor
of 30 over an hour and a half using a 0.44$''$ $\times$ 10$''$ slit.
Even though these different observations sample different wavelengths
and spatial regions, they all show the ejecta expanding and evolving
over tens of minutes to an hour and a half.

\begin{figure}[!h,!t,!b]\includegraphics[width=.9\linewidth,angle=90]{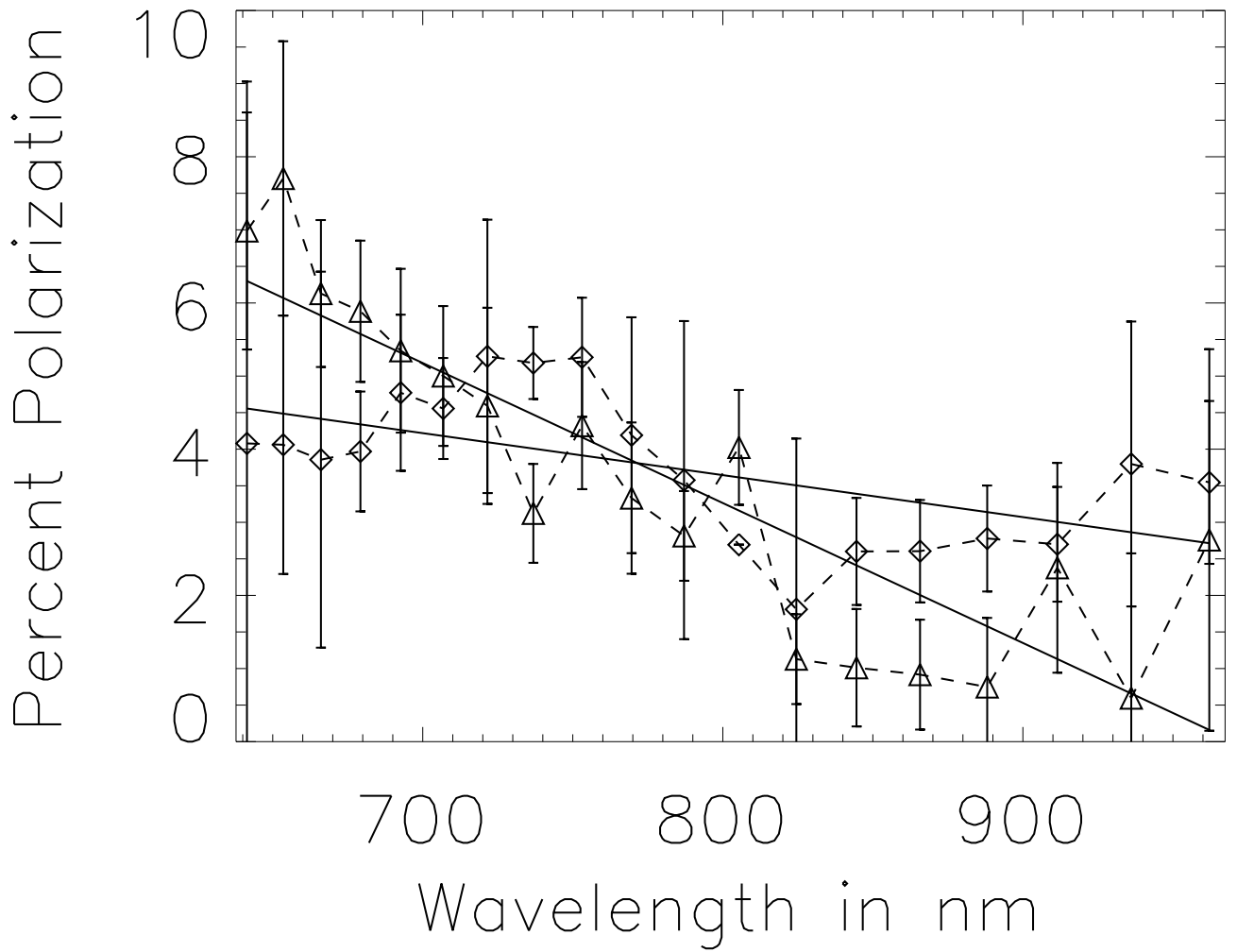}
\caption[tem_cper]{\label{fig-tem_cper}Degree of polarization for comet 9P/Tempel 1 -- Corrected for telescope
induced polarization shown with 3-$\sigma$ error bars.  The data was taken
on impact night from 6-7 UT (top curve at 900nm, shown with diamonds)
and 7-8 UT (bottom curve at 900nm, shown with triangles) at a phase angle of 40.9$^\circ$.  A linear fit
is plotted to guide the eye.  A single point at 805nm (order 11) in the
6-7 UT data set has been replaced by the average of neighbor points,
and has no error bars.  The polarization spectra from 6-7 UT curve had a
shallow negative slope of -0.9$\pm$0.2\%/ 10$^3$ \AA.
The 7-8UT curve had a slope of -2.3$\pm$0.3\% / 10$^3$ \AA.}
\end{figure}

In order to present high signal to noise measurements of the degree of
polarization, the spectra (1000 pixels per order) were rebinned 100:1
(giving a theoretical S/N$\sim$100 with 10 resolution elements per
order) before doing the polarization calculations described above. The
Stokes parameters $q$ and $u$ were then averaged to produce a single
measurement per order (1000:1). Each fractional difference measurement
($q$ or $u$) and the polarization calibrations (see the calibration paper
by Harrington {\it et al.} 2006) used 13000 pixels.  The error was calculated
($\sigma/\sqrt{N}$) from the standard deviation of the 10 measurements per
order.  The spectropolarimetry is plotted in Fig. \ref{fig-tem_cper}.
A linear fit to each measurement is shown to guide the eye and illustrate
the trends.  The 6-7 UT data set, started 8 minutes after impact, shows a
mildy anomalous blue-sloped degree of polarization of 4\% falling to 3\%
from 650 to 950nm.  In contrast, the 7-8UT data set, started 75 minutes
after impact, shows a more pronounced anomalous (blue) slope from 7\%
at 650nm to 2\% at 950nm.

Calibration of comet spectropolarimetry is performed by interpolating
the unpolarized standard star observations to the pointing of the
comet to create a telescope-induced polarization spectrum. A third-order
polynomial fit was performed to smooth out the noise in the interpolation.
This telescope polarization is then subtracted from the observed comet
polarization to calibrate the comet measurements.  The corrections were
typically 2-3\% with mild wavelength dependence ($\leq1$\%) and they
did not alter the slope-change seen in the comet 9P/Tempel 1 and C/2004
Machholz spectropolarimetry.

The polarization spectra from 6-7 UT (10-70 minutes after impact) had a
shallow negative slope of -0.9$\pm$0.2\%/ 10$^3$ \AA\ from 650 to 950nm,
which changed to -2.3$\pm$0.3\% / 10$^3$ \AA\ in the 7-8 UT spectra
(70-130 minutes after impact) (see Fig. \ref{fig-tem_cper}).  This is an
indication of the change in particle scattering properties.

\section{Comparison to Comet C/2004 Q2 Machholz}

\begin{figure}[!h,!t,!b]
\includegraphics[width=.9\linewidth,angle=90]{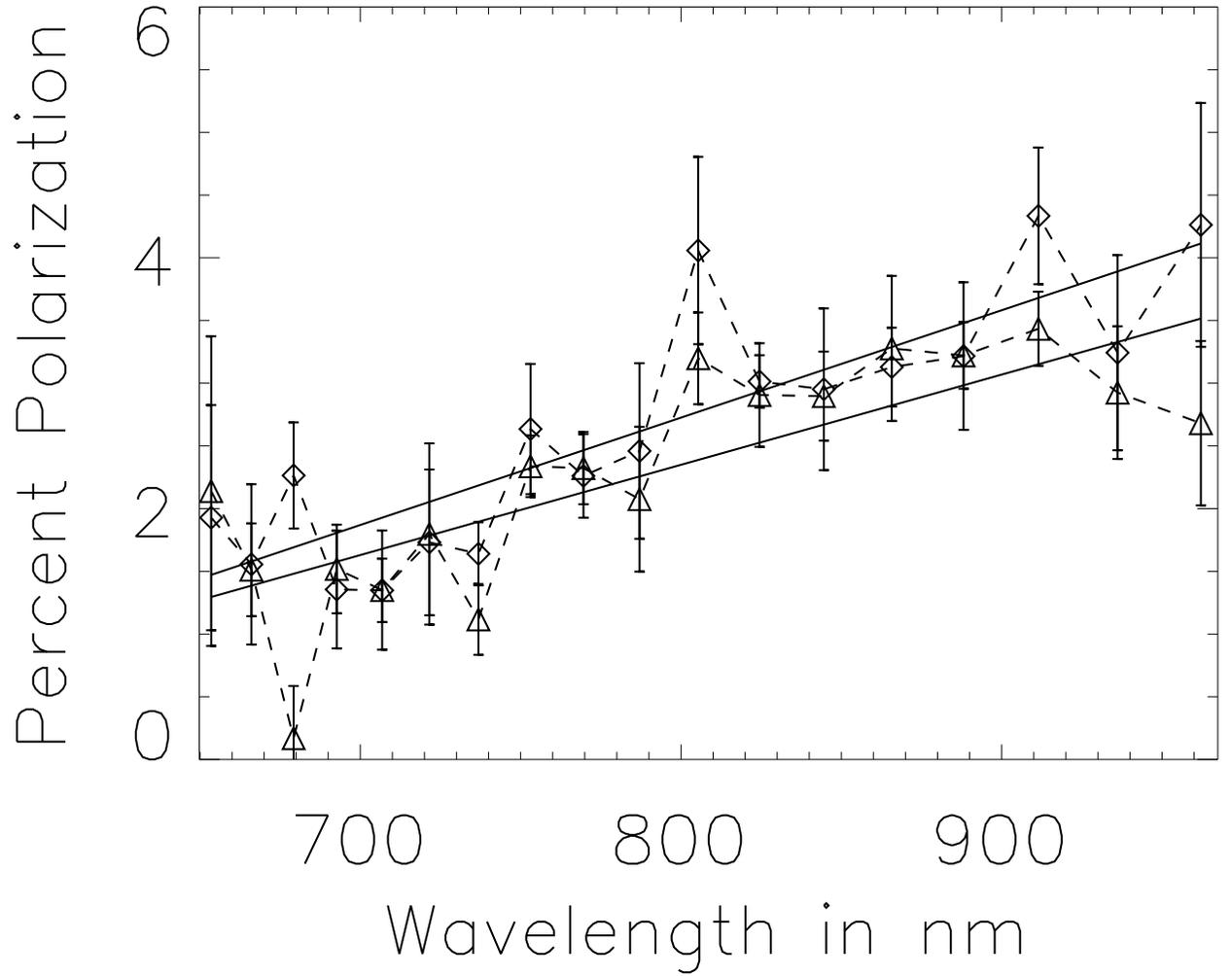}
\caption[mac_cper]{\label{fig-mac_cper}Degree of polarization for comet C/2004 Machholz at a phase angle of 30$^\circ$, corrected for telescope
induced polarization shown with 3-$\sigma$ error bars.  A linear fit is
lotted as a guide.  Both data sets follow each other very well and show
a typical red polarization slope gradually increasing with $\lambda$.  Slopes were +1.0 and +0.9 $\pm$0.1\% \ 10$^3$ \AA.
}

\end{figure}

Comet C/2004 Q2 Machholz was observed on November 27, 2004 at a phase
angle of $\alpha$=30$^\circ$ where most comets show a few percent
polarization.  Table~\ref{tab-macobs} summarizes the observations and
tabulates the change in pointing between images.  Two complete data sets
(8 images at 1200s) were reduced in the same way as the comet 9P/Tempel
1 data and the results are shown in Fig. \ref{fig-mac_cper}.  The 20
minute exposures had 40-180 DN per pixel, about double that of the comet
9P/Tempel 1 observations.  The measured degree of polarization did not
change very signifigantly between the two image sets (separated by 80
min) even though the azimuth changed by 20$^\circ$ in 80 minutes (a data
set).  This is in contrast with the much stronger change between sets
and the strong anomalous slope seen in comet 9P/Tempel 1 after impact.
The spectropolarimetry of comet C/2004 Machholz also shows a more typical
red comet polarization spectra and has values within the usual range
for comets at this phase angle (+1.0 and +0.9 $\pm$0.1\% / 10$^3$ \AA), giving us
confidence in our instrument and reduction techniques.

\section{Discussion and Conclusions}

The main result of our observations is an anomalous polarization spectrum with polarization decreasing with the wavelength (so called blue polarization spectrum). As mentioned above, such a behavior is not very typical for comets but has been seen before (Kolokolova {\it et al.} 2004, Kiselev {\it et al.} 2000).  Our observation of 4\% and 7\% polarization at 650nm is typical for comets at these wavelengths and phase angles, 4\% being somewhat low, but the 1\% and 3\% polarization at 950nm is not at all typical.  

Since a blue spectral gradient of polarization is not typical for comets, there are not many computer or laboratory simulations that have modeled it.  Blue polarization can result from larger particles or a predominance of transparent particles (silicates or ices).  For example, Kiselev {\it et al.} (2000) interpreted the blue polarimetric color of comet 21P/Giacobini-Zinner as evidence for either large dust particles or a predominance of carbonaceous materials.  Both these explanations arose from the knowledge that small particles in the Rayleigh scattering limit show red polarization and observations of the infrared spectra of this comet, reported by Hanner {\it et al.} (1992), which showed a very weak and broad silicate feature typical for dust with depletion of silicates or/and predominance of large particles.  However, the infrared observations of comet 9P/Tempel 1 right after the impact showed the situation opposite to the observed in comet 21P/Giacobini Zinner.  A strong and complex silicate feature developed by 1 hour after impact and decreased 1.8 hours after impact (Harker {\it et al.} 2005, Sugita {\it et. al.} 2005, Lisse {\it et al.} 2006). A relative lack of organics was also seen making us suspect that the silicates are responsible for the anomalous polarization slope.

The size of the silicate particles also evolved strongly and changed in the inner coma after the impact.  The pre-impact dust spectrum in the IR (5-35 $\mu$m) seen by Spitzer (Lisse {\it et al.} 2006) and by Gemini-North (8-13 $\mu$m) (Harker {\it et al.} 2005) was relatively featureless, likely due to the predominance of large dust aggregates in the coma, typical of other low activity comets.  Post-impact a large number of sharp emission features were seen both data sets.  The best fit size distribution reported for the Spitzer data showed that the dominant ejecta particle size fell between 1-5 $\mu$m, and that 0.1-10 $\mu$m particles were required to model the spectrum. Spectral models showed that both carbonaceous material and silicates were found, and that the total amount of carbonaceous material was about 20\% of the amount of silicates.  Harker {\it et al.} (2005) used a Hanner grain size distribution and found the peak in the size distribution to be 0.9$\mu$m pre-impact, 0.3$\mu$m one hour after impact, and 0.5$\mu$m 1.8 hours after impact.  These models showed that pre impact there was an absence of sub-$\mu$m silicates, and the spectrum was mineralogically dominated by larger (0.9$\mu$m) amorphous olivine with no carbon or crystalline silicates (see Harker {\it et. al.} 2005, table 1).  Post-impact their models showed an increase in the number of sub-$\mu$m silicates, as well as an increase in amorphous carbon, but an emission from relatively transparent Mg-rich crystalline olivine was also seen. They found that the silicate to carbon ratio roughly doubled between the first and second hour post-impact and that the crystalline to amorphous silicate ratio increased by a factor of 3.6.  They then suggest that this could result from smaller grains moving faster, size-sorting the ejecta cloud and leaving their later measurements to be dominated by larger, more crystalline grains (0.5$\mu$m vs 0.3$\mu$m).  

Harker {\it et al.} (2005) also speculate about the fragmentation of the aggregates further enhancing the spectral features.  The silicate grains are assumed to be coated by fine-grained amorphous carbon, as most IDP's are.  The sublimation and fragmentation process can disrupt this coating and free the silicate grains.  This is consistent with our measurements because larger, silicate-rich particles can cause blue polarimetric slopes.

Fern\'andez {\it et al.} (2006) reported a change in color of the dust in the near-IR post impact, becoming more blue during the first 14 minutes post impact, the change in color tracking the brightness increase, and becoming only marginally bluer after that.  They interpreted the color change as icy grains, or icy grain mantles being liberated in the impact event, which survived the observing period on the first night. There was direct observation of icy grains in the form of the 3 $\mu$m ice absorbtion from immediately after impact through lookback 46 min later (A'Hearn 2006 and references therein).  Schulz {\it et al.} (2006) reported evidence for icy grains in the inner 600km of the coma with the sublimation maximized 1.5 hours after impact, and then fading (presumably sublimating) thereafter.  Hodapp {\it et al.} (2006) also report evidence for the breakup of particles and the different spatial evolution of CN, [OI], and dust continuum flux in spectrophotometric measurements.  Thus, ice is also a possibility as a contribution to the transparency of the particles.

Multiple scattering depolarizes light in cometary dust and may be responsible for the low polarization just after impact.  Schleicher {\it et. al.} (2006) report an optically thick plume thinning to $\tau\sim0.4$ 20-25 minutes after impact.  The look-back images from the fly-by spacecraft also showed an optically thick ejecta plume after impact (A'Hearn {\it et al.} 2005).  Since the optical depth was a strong function of time, the influence of multiple scattering is assumed to change strongly as well.

We can explain our observations by a depolarization due to multiple scattering in the first hour and subsequent domination of larger and more transparent monomers (silicates) in the ejected dust aggregates of comet 9P/Tempel 1.  This is consistent with the infrared data on Deep Impact (Lisse {\it et al.} 2006, Harker {\it et al.} 2005, Sugita {\it et al.} 2005), which indicate a high amount of silicates in the DI dust whereas in situ data for comets 1P/Halley and 81P/Wild 2, typical red polarization comets, show that their dust contained two-thirds of carbonaceous materials and organics (Greenberg \& Li 1999, Kimura {\it et al.} 2006, Kissel {\it et al.} 2004). We can speculate that subsurface materials in comet 9P/Tempel 1 had more volatile organics and ices, as seen by other observers (Fern\'andez {\it et al.} 2006, Mumma {\it et al.} 2005, A'Hearn {\it et al.} 2005).  This is not surprising since on the surface refractory materials highly processed by radiation could survive. The volatile organics and ices quickly evaporated or decayed leaving the disrupted and fragmenting dust, rich in silicates to produce the observed blue polarization slope.  Some have suggested that smaller particles in the ejecta moved faster, size sorting the cloud, leaving larger and more crystalline silicates behind to cause the anomalous blue polarization slope (Harker {\it et al.} 2005).    Observations of comet Machholz showed typical behavior and give us confidence in our results and our instrument.

\section{Acknowledgements}

Support for this work was partially provided through University of Maryland and
University of Hawaii subcontract Z667702, which was awarded under prime
contract NASW-00004 from NASA, and also through an NSF grant.

\clearpage

\begin{table}[!h,!t,!b]
\begin{center}
\caption{HiVIS Comet 9P/Tempel 1 Observations \label{tab-obs}}
\begin{tabular}{cccccccccccc}
\hline
{\bf Date} & {\bf UT} & {\bf Exp [s]} & {\bf $\theta^{\dag}$} & {\bf Alt$^{\ddag}$} & {\bf Az$^{\ddag}$} &{\bf Date} & {\bf UT} & {\bf Exp [s]} & {\bf $\theta^{\dag}$} & {\bf Alt$^{\ddag}$} & {\bf Az$^{\ddag}$} \\
\hline
0630 & 06:40 &  1200 & 1 & 55 & 217  &0704 & 07:05 &   900 & 1 & 49  & 225 \\
0701 & 06:10 &  1200 & 1 & 58  & 205  &0704 & 07:25 &   900 & 2 & 46  & 230 \\
0701 & 06:30 &  1200 & 2 & 55  & 212  &0704 & 07:45 &   900 & 3 & 42  & 234 \\
0701 & 06:55 &  1200 & 3 & 52  & 223  &0704 & 08:00 &   900 & 4 & 39  & 237 \\
0701 & 07:20 &  1200 & 4 & 49  & 227  &0704 & 08:15 &  1800 & 1 & 36  & 240 \\
\hline
0703 & 06:10 &   600  & 1 & 57 & 206  &0705 & 06:35 &  1200 & 1 & 53 & 214 \\
0703 & 06:20 &  1200 & 1 & 55 & 208 &0705 & 06:55 &  1200 & 2 & 49 & 223  \\
0703 & 06:40 &  1200 & 2 & 53 & 217  &0705 & 07:15 &  1200 & 3 & 47 & 228  \\
0703 & 07:00 &   600  & 1 & 51 & 223  &0705 & 07:35 &  1200 & 4 & 44 & 231  \\
\hline
0704 & 06:00 &   900 & 1 & 58  & 202 &0706 & 05:55 &  1800 & 1 & 56 & 202  \\
0704 & 06:15 &   900 & 2 & 56  & 208 &0706 & 06:30 &  1800 & 2 & 53 & 214 \\
0704 & 06:30 &   900 & 3 & 54  & 214 &0706 & 07:00 &  1800 & 3 & 49 & 223 \\
0704 & 06:45 &   900 & 4 & 52  & 219 &0706 & 07:30 &  1800 & 4 & 43 & 231 \\
\hline
\end{tabular}
\end{center}
Notes:  $^{\dag}$ Waveplate orientation  $^{\ddag}$ Telescope altitude and
azimuth.  UT is the start time of the exposure.
\end{table}

\clearpage

\begin{table}[!h,!t,!b]
\large
\begin{center}
\caption{Comet C/2004 Machholz Observations -- 11-27-2004 \label{tab-macobs}}
\begin{tabular}{rcccc}
\hline
{\bf Start UT} & {\bf Exp [s]} & {\bf $\theta$$^{\dag}$} & {\bf Alt$^{\ddag}$} & {\bf Az$^{\ddag}$} \\
\hline
 8:15 & 1200 & 1 & 28 & 140 \\
 8:35 & 1200 & 2 & 31 & 144 \\
 9:00 & 1200 & 3 & 34 & 150 \\
 9:25 & 1200 & 4 & 37 & 156 \\
\hline
 9:45 & 1200 & 1 & 39 & 161 \\
10:05 & 1200 & 2 & 40 & 166 \\
10:25 & 1200 & 3 & 41 & 172 \\
10:45 & 1200 & 4 & 41 & 178 \\
\hline
\hline
\end{tabular}
\end{center}
Notes: $^{\dag}$Waveplate orientation $^{\ddag}$Telescope altitude and
azimuth.
\end{table}

\clearpage

\clearpage

\clearpage
	
\clearpage

\clearpage

\end{document}